\documentclass[showpacs,amsmath,amssymb,twocolumn,floatfix,prl]{revtex4}
\usepackage[dvips]{graphicx}
\input{epsf}
 
\def\cnot{\textrm{C\scshape not}}

\begin{document}

\title{Quantum computing of delocalization in small-world networks} 

\author{O. Giraud, B. Georgeot and D.L. Shepelyansky} 
\affiliation{Laboratoire de Physique Th\'eorique, UMR 5152 du CNRS, 
Universit\'e Paul Sabatier, 31062 Toulouse Cedex 4, France}

\date{March 23, 2005} 

\begin{abstract}
We study a quantum small-world network with disorder
and show that the system exhibits
a delocalization transition. 
A quantum algorithm is built up
which simulates the evolution operator
of the model in a polynomial number of gates for exponential number of
vertices in the network.  The total computational gain
is shown to depend on the parameters of the network
and a larger than quadratic speed-up can be reached.
 We also investigate the robustness of 
the algorithm in presence of imperfections.
\end{abstract}
\maketitle

Recently, much attention has been attracted 
to the study of small-world networks \cite{smallworld}.
They have been shown to describe social and biological 
networks, Internet connections,
airline flights and other complex networks.
 In such systems, it is possible to
go from a given point to any other through only a small number 
of links.
Well-established classical 
models have been proposed and analyzed by statistical methods.
The study of quantum networks with the same property
has started only recently, showing that these
systems present interesting features related to quantum transport,
delocalization \cite{chinois,como2001} 
and fast diffusion \cite{diffusion}.

In parallel, the development of quantum information and computation has become
more and more important \cite{Nielsen}.  In particular, the
study of quantum computers has shown that they can solve
certain problems much more efficiently than any classical
device.  Celebrated quantum algorithms have been built for the factorization of
large numbers with exponential efficiency
\cite{shor}, and for search in 
an unstructured database with a quadratic speed-up
\cite{grover}.
As first envisioned by Feynman in the 1980's, 
the simulation of complex quantum systems has also been shown to 
be more efficient on a quantum computer \cite{Nielsen}.

Here we study a quantum
small-world network with disorder.  We demonstrate the existence of
a delocalization transition
and investigate its dependence on disorder strength,
number of links and system size.  We then build a
quantum algorithm to simulate such a network on a quantum computer,
and show that its efficiency significantly overcomes 
classical computations. The algorithm is 
robust with respect to errors.

We consider a circular graph with $N=2^{n_r}$ vertices. Each vertex is 
linked with its two nearest-neighbors. To this graph,  $pN$ shortcut
links (connecting $2pN$ vertices)
 are added between random
pairs of vertices (see an example in the inset of Fig.1)
\cite{footnote}.
A quantized version of this system with on-site disorder can be
described by the $N \times N$ Hamiltonian matrix
$H=H_0+H_1+H_2$.  The first two terms give a one-dimensional
tight-binding Anderson model well-known in solid state physics \cite{mirlin}.
The diagonal matrix with entries $(H_0)_{ij}=\epsilon_i \delta_{i,j}$ 
describes on-site disorder; 
$\delta_{i,j}$ denote Kronecker symbols, and $\epsilon_i$ are 
independent random numbers whose distribution
is a Gaussian with zero mean and width $W$ 
(the Gaussian is truncated at large values). The matrix
$(H_1)_{ij}=V (\delta_{i, j+1}+\delta_{i+1, j})$
describes the links between nearest-neighbors, and  
$(H_2)_{ij}=\sum_{k=1}^{M}V (\delta_{i,i_k}\delta_{j, j_k}+
\delta_{i,j_k}\delta_{j, i_k})$ the
shortcuts which make the graph of small-world type, 
where $\{i_k,j_k\}$ are the pairs of vertices connected by random  
links, and $V=1$ is the hopping matrix element. 

\begin{figure}[h]
\begin{center} 
\includegraphics[width=.85\linewidth]{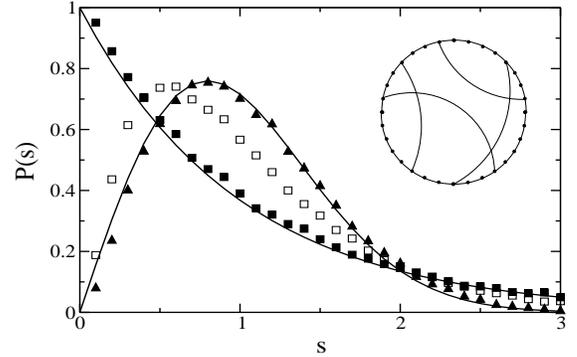}
\end{center} 
\caption{Level spacing statistics for $H$ at $n_r=14$,
$p=1/32$, for three values of the disorder: $W=0.5$ (triangles), 
$1.3$ (empty squares) and $3$ (full squares). 
The solid curves correspond to the Poisson 
distribution $P(s)=e^{-s}$ and to the Wigner-Dyson distribution 
$P(s)=(\pi s/2)e^{-\pi s^2/4}$. Number of disorder realizations 
(position of shortcut links and on-site disorder) is
$N_D=10$. Only the central half of the 
eigenvalues is taken into account. Inset:
a realization of small-world network with $N=32$ and $p=1/8$.} 
\label{fig1}
\end{figure}

When $p=0$, the system reduces to the one-dimensional Anderson
model, for which
all states are known to be
localized. For small disorder, the localization length $l$
varies as $l \propto 1/W^2$ \cite{mirlin}.
The additional presence of shortcut links may induce 
delocalization. This can be checked
through spectral statistics.  Indeed, for localized systems, the
eigenvalues are distributed according to the Poisson distribution,
provided the localization length is smaller than the system size.
On the contrary, in the delocalized phase
the eigenvalues follow the Wigner-Dyson distribution corresponding
to Random Matrix Theory, which generally characterizes
quantum chaotic systems  and ergodic wavefunctions \cite{mirlin}.
Our numerical diagonalization of $H$ at fixed $p$ shows
 a transition from Poisson to
Wigner distribution as $W$ decreases. A typical example
is shown in Fig.1 at $p=1/32$ and $W= 3$ (localized phase),
$W= 1.3$ (intermediate statistics), $W=0.5$ (delocalized phase). 
This indicates that a delocalization transition takes place
in this system.

\begin{figure}
\begin{center} 
\includegraphics[width=.65\linewidth, angle=-90]{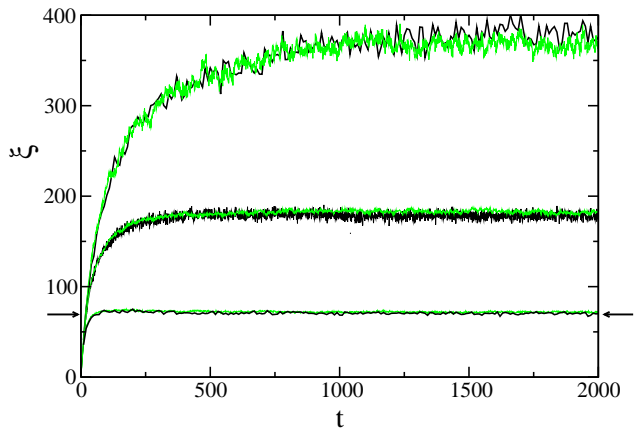}
\end{center} 
\caption{(color online) Evolution of the IPR  $\xi$ with time $t$, for 
$W=0.5$ and $p=1/32$. Initial state is localized on one vertex.
Curves correspond (from bottom to top) to 
$n_r=8,10,12$. Each curve is shown as obtained by exact evolution
(black lines), with $N_D=80$, 
and by simulation by quantum gates (see text)(green/gray lines)
with $N_D=100$ and $\Delta t=0.03$.
The arrows indicate the IPR at $t=2000$ in the absence of
shortcut links $p=0$ (exact evolution and simulation by quantum
gates yield the same result, data not shown).}
\label{IPR05} 
\end{figure} 

\begin{figure}
\begin{center} 
\vspace{0.5cm}
\includegraphics[width=.65\linewidth, angle=-90]{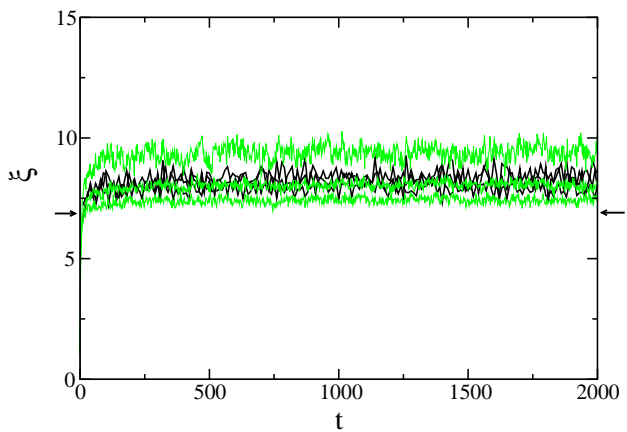} 
\end{center} 
\caption{(color online) Same as Fig.\ref{IPR05} with $W=3$.}
\label{IPR3}
\end{figure} 

\begin{figure}[h] 
\vspace{0.7cm}
\begin{center} 
\includegraphics[width=.65\linewidth, angle=-90]{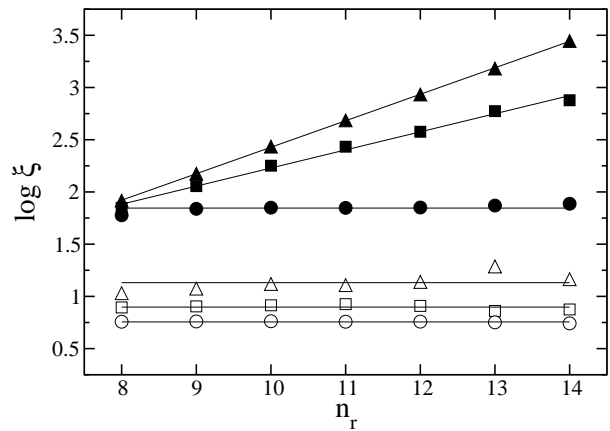}
\end{center} 
\caption{IPR at time $t=2000$ as a function of 
$n_r$ for $W=0.5$ (full symbols) and $W=3$ (empty symbols), for 
$p=1/16$ (triangles), $p=1/32$ (squares)
and $p=0$ (circles), with $20 \leq N_D \leq 160 $.  Initial state is localized
on one vertex.
Straight lines correspond from top to bottom 
to $\xi=0.78 N^{0.84}$, $\xi=3.13 N^{0.58}$, 
$\xi=70.15$, $\xi=13.55$, $\xi=7.91$, $\xi=5.69$,
with $N=2^{n_r}$.
Logarithm is decimal.}
\label{logs}
\end{figure}

The localization properties of this quantum system can 
be analyzed more precisely
through the Inverse Participation Ratio (IPR), defined by
$\xi = \sum_i|\Psi_i|^2/\sum_i|\Psi_i|^4$ for a wavefunction 
$|\Psi \rangle=\sum_i\Psi_i |i\rangle $.
It gives the number of vertices supporting the wavefunction  
($\xi=1$ for a state localized on a single vertex, and $\xi=N$ for 
a state uniformly spread over $N$ vertices).
In Fig.\ref{IPR05} and Fig.\ref{IPR3}, we display the time evolution of the
IPR for a wave packet initially localized on one vertex.
For $W=0.5$, the saturation value grows with $N$ in the presence of
shortcut links, indicating that the wavefunction is no longer localized.
On the contrary, for $W=3$, the saturation value remains close to its
value in the absence of links and does not change significantly with $N$,
implying that the system is still localized.
In a more quantitative way, Fig.\ref{logs} presents the saturation 
value of the IPR as a function of $n_r$ for different values of $p$.
The data confirm that at $W=3$ the system remains localized.
On the contrary, a clear delocalization is visible
in the presence of shortcut links for $W=0.5$.   
The data are in good agreement with the law $\xi \propto N^{\alpha}$,
with $\alpha \approx 0.58$ for $p=1/32$ and $\alpha \approx 0.84$ for $p=1/16$
(the maximal value $\alpha=1$ is obtained at $p=1/2$, data not shown).  
This shows that the delocalization transition for $p=1/16$ and $p=1/32$
takes place approximately at $W \approx 1$.
In the limit of weak disorder $W \ll 1$, the transition is expected
to take place at 
smaller values of $p \propto W^2$ \cite{como2001}.

This system can be simulated on a quantum computer, using $O((\log N)^2)$
quantum gates for a network of $N=2^{n_r}$ vertices, and $n_q=3n_r+3$ qubits.
We start from an initial wave packet encoded on the quantum registers.
For example, the initial one-vertex states used in Figs.\ref{IPR05}-\ref{logs}
can be constructed efficiently
from a state localized in the ground state of the quantum computer
by at most $n_r$ single-qubit flips.
Our quantum algorithm performs the evolution of the wave packet 
by slicing the propagator $\exp(i H t)$, using the relation 
$e^{i(H_0+H_1+H_2)\Delta t}=
e^{i H_0 \frac{\Delta t}{2}}
e^{i H_1 \frac{\Delta t}{2}} e^{i H_2\Delta t}e^{i H_1 \frac{\Delta t}{2}}
e^{i H_0 \frac{\Delta t}{2}}+O(\Delta t^3)$ for a short period of time 
$\Delta t$ (see e.g. \cite{Nielsen,pomerans}). Each unitary operator is then
simulated by quantum gates.
We use in particular rotations on the $j$-th qubit by an angle $\phi/2$:
$R_j(\phi)=\exp(i\phi\sigma_j^{z}/2)$ ($\sigma^z$ being a Pauli matrix);
controlled-not operations $\cnot_{i,j}$, that is
bit-flip on the $j$-th qubit conditioned by the $i$-th qubit;
multi-controlled rotations
$C_{i_1, ..., i_{\mu},\eta_{1}, ...,\eta_{\mu}, j}(\theta)$, that is
rotations by an angle $\theta$ on the $j$-th qubit if and only if the qubits 
$i_k$ takes the value $\eta_k\in\{0,1\}$ for $1\leq k \leq \mu$.

$\bullet$ The transformation 
$|i\rangle\rightarrow e^{i H_0 \Delta t} |i\rangle$ consists in
multiplying each basis state $|i\rangle$ by a Gaussian random phase 
$\exp(i \epsilon_i \Delta t)$. For some integer $n_s$, and
$\sigma=W\Delta t \sqrt{\frac{3}{n_r+n_s}}$, let us choose 
randomly $n_r+n_s$ angles $\phi_k$, $1\leq k\leq n_r$, and $\phi'_k$,  
$1\leq k\leq n_s$, independent and uniformly distributed in 
$[-\sigma/2, \sigma/2]$. Each $\epsilon_i \Delta t$ is replaced by a
random variable $\pm \phi_1\pm\phi_2\pm\cdots\pm\phi'_{s-1}\pm\phi'_{s}$,
which for large $n_s$ tends to a Gaussian random variable of width 
$W\Delta t$. This can be simulated by applying the operator 
$\prod_{k=n_s}^{1} \cnot_{i_{k}, j_{k}} 
\prod_{k=1}^{n_s}\left(R_{j_k}(\phi'_k) \cnot_{i_k, j_k}\right)
\prod_{k=1}^{n_r}R_k(\phi_k)$
for some value of $n_s$. 
The $i_k$ and $j_k$ are chosen randomly between $0$ and $n_r-1$.
This step requires $(3n_s+n_r)$ gates.

$\bullet$ To perform the transformation 
$|i\rangle\rightarrow e^{i H_1 \Delta t} |i\rangle$,
we first apply a quantum Fourier transform (QFT)
to turn it
into the diagonal transformation
$|k\rangle\rightarrow\exp\left(2i\Delta t\cos\frac{2\pi k}{N}\right)
|k\rangle$.
Following \cite{pomerans}, we introduce the operator 
$R_{\gamma}(\bar{\theta})=H S^1 H e^{-i\frac{\gamma}{2}\sigma_1^z} 
H S^{-2} H e^{-i\frac{\gamma}{2}\sigma_1^z}H S^1$ with 
$S^m=\prod_{j=2}^{n_r}C_{1,j}(\pi m/ 2^{j-1})$. 
It can be shown that 
$e^{-i\gamma\cos\theta}=R_{\gamma/2}(\bar{\theta})R_{\gamma/2}(-\bar{\theta})
+O(\gamma^3)$ for small $\gamma$, with $\bar{\theta}=\theta-\pi a_1$ 
if $\theta/2\pi=0.a_1 a_2\ldots a_n$. The diagonal operator can thus be 
approximated by $\exp\left(2i\Delta t\cos\frac{2\pi k}{N}\right)=
\left(R_{\gamma/2}(\bar{\theta})R_{\gamma/2}(-\bar{\theta})\right)^L
+O(L\gamma^3)$, with $L$ an integer and $\gamma$ a small parameter
chosen such that $L\gamma=-2i\Delta t$. We then perform an inverse QFT.
The two QFT require $n_r(n_r +1)$ gates, and 
the simulation of the diagonal term requires $2L(5+3n_r)$ gates.

$\bullet$ The transformation 
$|i\rangle\rightarrow e^{i H_2 \Delta t} |i\rangle$
acts on the subspace spanned by $|i_k\rangle$ and 
$|j_k\rangle$, where  $i_k$ and $j_k$ are linked by a shortcut link,
through the $2\times 2$ submatrix 
$e^{i\Delta t\sigma^x}$.  
Let us first assume that $p$ is of the form $p=1/2^{\rho}$. 
For $\mu=\rho-1$, the operator
$C_{i_1, ..., i_{\mu},\eta_{1}, ...,\eta_{\mu},j}(\theta)$
acts on the $2^{n_r-\mu}$ basis vectors whose qubits $i_k$, $1\leq k \leq\mu$, 
are respectively equal to $\eta_k$: it corresponds to the creation
of $2^{n_r-\mu-1}=2^{n_r-\rho}=pN$ links.
In order to have less regular shortcut links, we first perform a permutation 
on the vertices. To do this, we randomly choose $n_p$ integers $a_k$ and 
$b_k$, for some integer $n_p$. It is better to take the $a_k$ 
in $[0.2 N , 0.8 N]$ and odd. Then we define the operators 
$U_k|i\rangle=|(a_k i+b_k) \mod N\rangle$ and the inverse operators
$V_k|i\rangle=|a_k^{-1} (i-b_k) \mod N\rangle$.
A permutation can be simulated by the sequence of gates 
$P=\prod_{k=1}^{n_p}U_k\ \cnot_{i_k, j_k}$ 
where the $i_k$ and $j_k$ are chosen randomly. 
Application of the permutation $P$, followed by a multi-controlled rotation
$C_{i_1, ..., i_{\mu}, \eta_{1}, ...,\eta_{\mu},j}(\Delta t)$ and
$P^{-1}$, gives $e^{iH_2 \Delta t}$. The $i_k$ and $\eta_k$ 
in the controlled rotation are also chosen randomly. 
In the general case, where $p\neq 1/2^{\rho}$, 
we expand $pN$ in base 2, such that 
$pN=\sum 2^{p_k}$. Then we replace the multi-controlled gate in the above
description by a multi-controlled gate for each $p_k$ appearing
in the decomposition of $pN$.  This gives a sequence of gates 
$C_{i^{(k)}_1, ..., i^{(k)}_{\mu_k}, \eta^{(k)}_{1}, ...,\eta^{(k)}_{\mu_k}, j^{(k)}}(\Delta t)$, where $\mu_k=n_r-p_k-1$, and the 
$i^{(k)}_{k'}$, $j^{(k)}$ and $\eta^{(k)}_{k'}$ are chosen randomly. 
Each operator $U_k$ consists of a multiplication and an addition modulo $N$,
which can be performed using $(2n_r+3)$ ancilla qubits and $O(n_r^2)$ 
quantum gates \cite{arithm1}.
Each multi-controlled gate can be performed by $O(n_r^2)$ Toffoli, $\cnot$ and 
single qubit gates \cite{arithm2}.

In total, the simulation of a network of $N=2^{n_r}$ vertices
for one unit of time
with fixed parameters $\Delta t$, $n_s$, $L$ and $n_p$ can be done by 
this method with $O(n_r^2)$ quantum operations and $3n_r+3$ 
qubits. Classically, a similar method
can only be implemented in $O(N)$ operations at best.
The quantum simulation is therefore exponentially faster.
  This remains the case even if the parameters
$n_s$ and $n_p$ are allowed to grow linearly with $n_r$ to improve accuracy
(the cost becomes $O(n_r^3)$ quantum gates).  

The algorithm simulates the small-world network efficiently
but at the cost of several approximations.  
In order to check its convergence and accuracy, we implemented it
on a (classical) computer. In Figs.\ref{IPR05},\ref{IPR3}, we display
the result of this computation for the parameters $\Delta t =0.03$,
$n_s = 30 n_r$, $L=10$, and $n_p =3 n_r$ alongside the exact evolution,
showing that the algorithm is quite accurate for these values, and enables to 
monitor precisely the delocalization transition with good accuracy.  
The computation 
accuracy is not very sensitive to fixed values of 
$L$ and $\Delta t$: the total size $N$
can be changed by orders of magnitude (factor of $64$ in our case) without
modification of these parameters.

To estimate the total complexity of the algorithm, we should take
into account the number of quantum measurements and the number
of iterations of the map.  In order to see the delocalization transition, 
it is sufficient to estimate the spreading of the
wavefunction, which can be done by a constant number of quantum measurements
\cite{loclength}.
Still, the initial wave packet should have enough time
to spread in order for the localization length to be estimated.
For the parameters of Fig.\ref{IPR05}, we determined the time $\tau$ needed 
for the IPR to reach half of its maximal value.  
In the delocalized phase for $p < 1/2$,
our numerical results give the scaling $\tau \propto N^\beta$ with
$\beta \approx 0.83$ ($p=1/16$) and $\beta \approx 0.69$ ($p=1/32$)
(data not shown).  This means that the total cost of the 
quantum algorithm will scale as $O(N^\beta)$, compared to 
$O(N^{\beta +1})$ for the classical one (dropping logarithmic factors).
This implies a better than quadratic gain for the quantum computation,
but no exponential gain.  
In contrast, for $1/2 < p \leq 2$ we find that $\tau \approx  \log N$ 
(data not shown) \cite{footnote2}.
In this case, the algorithm
may reach exponential efficiency and enable to perform 
precise studies of this percolation-like transition for very
large values of $N$. The exact algorithm complexity depends on
the properties of the phase transition near critical $W$ value.

\begin{figure}[h] 
\vspace{0.5cm}
\begin{center} 
\includegraphics[width=.65\linewidth, angle=-90]{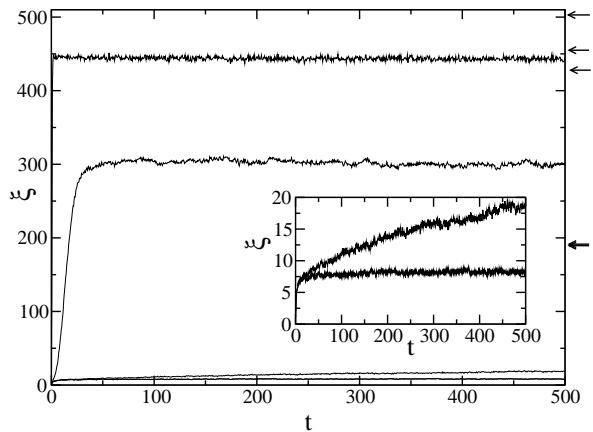}
\end{center} 
\caption{Evolution of the IPR with time in 
the presence of static errors, for $n_r=10$, $W=3$ and $p=1/32$, with
$N_D=100$ (static errors are the same for all realizations). Initial 
state is localized on one vertex, $\Delta t=0.03$.
From top 
to bottom: $\epsilon=10^{-4}$, $10^{-5}$, $10^{-6}$, $10^{-7}$ and $0$. 
The arrows mark $\xi$ at $t=500$ 
for $W=0.5$ and same other parameters, 
with from top to bottom $\epsilon=10^{-5}$, $10^{-4}$, 
$10^{-6}$, $10^{-7}$ and $0$. 
Inset shows the three lowest curves on a different scale. 
Data from $\epsilon=10^{-7}$ 
are indistinguishable from $\epsilon=0$.}
\label{errors}
\end{figure} 

These results show that a perfect quantum computer gives a 
significant gain in the simulation of quantum small-world networks.
However, realistic quantum computers are prone to errors and imperfections.
It is therefore important to test the resilience of the algorithm
to such effects.  In Fig.\ref{errors} we show the result of numerical
simulations of the algorithm in presence of errors. The error model
chosen corresponds to static imperfections. These errors can exist
independently of the coupling with the external world, and have parametrically
larger effects than random noise in the gates \cite{qchaos}.
Between each gate the system 
evolves through the additional Hamiltonian
$H_E = \sum_{i} \delta_i \sigma_{i}^z + \sum_{i} J_{i} 
\sigma_{i}^x \sigma_{i+1}^x$,
where the second
sum runs
over nearest-neighbor qubit pairs on a circular chain.
The $\delta_i$ are randomly and uniformly distributed in the interval 
$[-\delta /2, \delta/2 ]$. 
The couplings $J_{i}$ represent the residual static interaction 
between qubits and are chosen 
randomly and uniformly in the interval $[-J,J]$.
We suppose that each gate in the quantum algorithm is instantaneous
and separated by a time $\tau_g$ during which $H_E$ acts. We take 
one single rescaled parameter $\varepsilon$ which describes the 
amplitude of these static errors, with $\varepsilon=\delta\tau_g=J\tau_g$.
In the numerical simulations, to save computational time
we took the part of the algorithm which
generates the random shortcut links as exact, all other parts being performed 
with errors.
The results displayed in Fig.\ref{errors} show that with moderate levels
of imperfections ($\varepsilon \approx 10^{-7}$) 
the simulation of the small-world 
network is very close to the exact computation, in absence of any quantum
error correction.

In conclusion, we have shown that quantum disordered small-world networks,
which display a delocalization transition,
can be simulated more efficiently on quantum computers than on classical 
ones.  The algorithm can be performed accurately on realistic 
few-qubit quantum computers in presence of moderate error strength.

We thank A. Pomeransky and O. Zhirov for
helpful discussions.
 We thank the IDRIS in Orsay and CalMiP in Toulouse for 
access to their supercomputers.
This work was supported in part by the 
project EDIQIP of the IST-FET program of the EC.


\begin{thebibliography}{99}
\bibitem{footnote} Shortcuts do not connect nearest-neighbors.
\bibitem{footnote2}We note that in \cite{diffusion} a logarithmic
law for $\tau$ was also found in another type of quantum
small-world network
without on-site disorder.
\end{thebibliography}
\end{document}